\documentclass[runningheads,a4paper]{llncs}

\usepackage{amssymb}
\usepackage{graphicx}
\usepackage{datetime}

\usepackage{url}
\newcommand{\keywords}[1]{\par\addvspace\baselineskip
\noindent\keywordname\enspace\ignorespaces#1}
\usepackage[utf8]{inputenc}
\usepackage[english]{babel}
\usepackage{indentfirst}

\usepackage{cite}

\begin{document} 

\mainmatter  

\title{Neuromorphic Electronic Systems for Reservoir Computing}

\titlerunning{Neuromorphic Electronic Systems for Reservoir Computing}

\author{Fatemeh Hadaeghi}

\institute{Department of Computer Science and Electrical Engineering, Jacobs University Bremen,
28759 Bremen, Germany\\
Institute of Computational Neuroscience, University Medical Center Hamburg-Eppendorf (UKE),
20251 Hamburg, Germany\\
\email{f.hadaeghi@uke.de}}
%
\authorrunning{Neuromorphic Electronic Systems for Reservoir Computing}
\maketitle
%
%





\makebox[\linewidth]{\small 29 July 2020~~~~~~~~}
%
%
%

\begin{abstract}
This chapter provides a comprehensive survey of the researches and motivations for hardware implementation of reservoir computing (RC) on neuromorphic electronic systems. Due to its computational efficiency 
and the fact that training amounts to a simple linear
regression, both spiking and non-spiking implementations of reservoir computing on neuromorphic hardware have been developed. Here, a review of these experimental studies is provided to illustrate the progress in this area and to address the technical challenges which arise from this specific hardware implementation. Moreover, to deal with challenges of computation on such unconventional substrates, several lines of potential solutions are presented based on advances in other computational approaches in machine learning.

\keywords{Analog Microchips, FPGA, Memristors, Neuromorphic Architectures, Reservoir Computing}
\end{abstract}

\section{Introduction} 
The term ``neuromorphic computing" refers to a variety of brain-inspired computers, architectures, devices, and models that are used in the endeavor to mimic biological neural networks \cite{mead2012analog}. In contrast to von Neumann architectures, biologically inspired neuromorphic computing systems are promising for being highly connected and parallel, incorporating learning and adaptation, collocating memory and processing and requiring low-power. By creating parallel arrays of connected synthetic neurons that are asynchronous, real-time, and data- or event-driven, neuromorphic devices offer an expedient substrate to model neuroscience theories as well as implementing computational paradigms to solve challenging machine learning problems.

The accustomed growth rates of digital computing performance levels (Moore's Law) is showing signs of flattening out \cite{kish2002end}. Furthermore, the exploding energy demand of digital computing devices and algorithms is approaching the limits of what is socially and environmentally tolerable \cite{dreslinski2010near}.  Neuromorphic technologies suggest escape routes from both predicaments due to their potentials for unclocked parallelism and minimal energy consumption of spiking dynamics. Moreover, neuromorphic systems have also received increased attention due to their scalability and small device footprint \cite{schuman2017survey}.

Significant keys to such advancements are remarkable progress in material science and nanotechnology, low-voltage analog CMOS design techniques and theoretical and computational neuroscience. At the device level, using the new materials and nanotechnologies for building extremely compact and low-power solid-state nanoscale devices, has paved the way towards on-chip synapses with characteristic properties observed in biological synapses. For instance, ``memory resistor" or the memristor, a non-linear nanoscale electronic element with volatile and non-volatile modes, is ubiquitously used in neuromorphic circuits to store multiple bits of information and to emulate dynamic weights with intrinsic plasticity features (e.g., spike time dependent plasticity (STDP)) \cite{yang2013memristive,moon2019temporal}. It has been argued that hybrid memristor-CMOS neuromorphic circuit may represent a proper building block for implementing biological-inspired probabilistic/stochastic/approximate computing paradigms that are robust to memristor device variability and fault-tolerant by design \cite{indiveri2013integration,adam20173,jo2010nanoscale}. Similarly, conductive-bridging
RAM (CBRAM) \cite{suri2012cbram}, which is a non-volatile memory technology, atomic switches \cite{aono2010atomic} -- nanodevices implemented
using metal-oxide based memristors or memristive
materials -- and tin oxide nanoparticles \cite{le2020electroformed} have also been fabricated to implement both short-term plasticity (STP) and long-term plasticity (LTP) in neuromorphic systems \cite{avizienis2012neuromorphic,sillin2013theoretical}. Both atomic switches and CBRAM have nano dimensions, are fast, and consume low energy \cite{schuman2017survey}. Spike- time-dependent-depression and -potentiation observed in biological synapses have also been emulated by phase change memory (PCM) elements in hybrid neuromorphic architectures where CMOS ``neuronal" circuits are integrated with nanoscale ``synaptic" devices. PCM elements are commonly used to achieve high density and scalability; they are compatible with CMOS circuits and show a good endurance \cite{suri2012physical,ambrogio2016unsupervised}. Programmable metallization cells (PMCs) \cite{yu2010modeling} and oxide-resistive memory (OXRAM) \cite{yu2011electronic} are another types of resistive memory technologies that have been demonstrated to present STDP-like characteristics. Beyond the CMOS technologies
for neuromorphic computing, spintronic devices and optical (photonic) components have also been considered for neuromorphic implementation \cite{schuman2017survey,tanaka2019recent}.

At the hardware level, a transition away from purely digital systems to mixed analog/digital implementation to purely analog, unclocked, spiking neuromorphic microchips has led to the emergence of more biological-like models of neurons and synapses together with a collection of more biologically plausible adaptation and learning mechanisms. Digital systems are usually synchronous or clock-based, rely on Boolean logic-based gates and discrete values for computation and tend to need more power. Analog systems, in contrast, tend to rely more on the continuous values and inherent physical characteristics of electronic devices for computation and more closely resemble the biological brain which takes the advantages of physical properties rather than Boolean logic for computation \cite{indiveri2015memory}. Analog systems, however, are significantly vulnerable to various types of thermal and electrical noise and artifacts. It is, therefore, argued that only computational paradigms that are robust to noise and faults may be proper candidates for analog implementation. Hence, among the wide diversity of computational models, a leading role is emerging for
mixed analog/digital or purely analog implementation of artificial neural networks (ANNs). Two main trends in this arena are the exploit of low-energy, spiking neural dynamics for ``deep learning'' solutions \cite{merolla2014million}, and
``reservoir computing'' methods \cite{jaeger2004harnessing}.

At the network and system level, new discoveries in neuroscience research are being gradually incorporated into hardware models and mathematical theory frameworks are being developed to guide the algorithms and hardware developments. A constructive paradigm shift has also occurred from strict structural replication of neuronal systems towards the hardware implementation of systemic/functional models of the biological brain \cite{james2017historical}. Consequently, over the past decade, a wide variety of model types have been implemented in electronic multi-neuron computing
platforms to solve pattern recognition
and machine learning tasks \cite{indiveri2015memory,misra2010artificial}.

As mentioned above, among these computational frameworks, reservoir computing (RC) has been variously considered as a strategy to implement useful computations on such unconventional hardware platforms. An RC architecture comprises three major parts: the \emph{input layer} feeds 
the input signal into a random, large, fixed recurrent neural network that constitutes the \emph{reservoir}, from which the neurons in the \emph{output layer} read out a desired output signal. In contrast to traditional (and ``deep'') recurrent neural networks (RNNs) training methods, the input-to-reservoir and the recurrent reservoir-to-reservoir weights in an RC system are left unchanged after a random initialization, and only the reservoir-to-output weights are optimized during training.
Within computational neuroscience, RC is best known as \emph{liquid state machines} (LSMs) \cite{maass2002real}, whereas the approach is known as \emph{echo state networks} (ESNs) \cite{jaeger2004harnessing} in machine learning. Reflecting the different objectives in these fields, LSM models are typically built around more or less detailed, spiking neuron models with biologically plausible parametrizations, while ESNs mostly use highly abstracted rate models for its neurons. Due to its computational efficiency, simplicity and lenient requirements, both spiking and non-spiking implementations of reservoir computing on neuromorphic hardware exist.
However, proper solutions are still lacking to address a variety of technological and information processing problems. For instance, regarding the choice of hardware, variations of stochasticity due to device mismatch, temporal drift, and aging must be taken into account; in case of exploiting the spiking neurons, an appropriate encoding scheme needs to be developed to transform the input signal into spike trains; the speed of physical processes may require further adjustment to achieve on-line learning and real-time information processing; depending on the available technology, compatible local learning rules must be developed for on-chip learning.

Here, a review of recent experimental studies is provided to illustrate the progress in neuromorphic electronic systems for RC and to address the above-mentioned technical challenges which arise from such hardware implementations. Moreover, to deal with challenges of computation on such unconventional substrate, several lines of potential solutions are presented based on advances in other computational approaches in machine learning.
In the remaining part of this chapter, we present an overview of the current approaches to implement reservoir computing model on digital neuromorphic processors, purely analog neuromorphic microchips and mixed digital/analog neuromorphic systems. Since the neuromorphic computing attempts to implement more biological-like models of neurons and synapses, spiking implementations of RC will be highlighted. 

\section{RC on Digital Neuromorphic Processors}
Field programmable gate arrays (FPGAs) and application specific integrated circuit (ASIC) chips -- two categories of digital systems -- have been very commonly utilized for neuromorphic implementations. To facilitate the application of computational frameworks in both embedded and stand-alone systems, FPGAs and ASICs offer considerable flexibility, reconfigurability, robustness, and fast prototyping \cite{wang2017energy}.

\begin{figure} [tp]
	\begin{centering}
		\includegraphics[width=1.0\columnwidth]{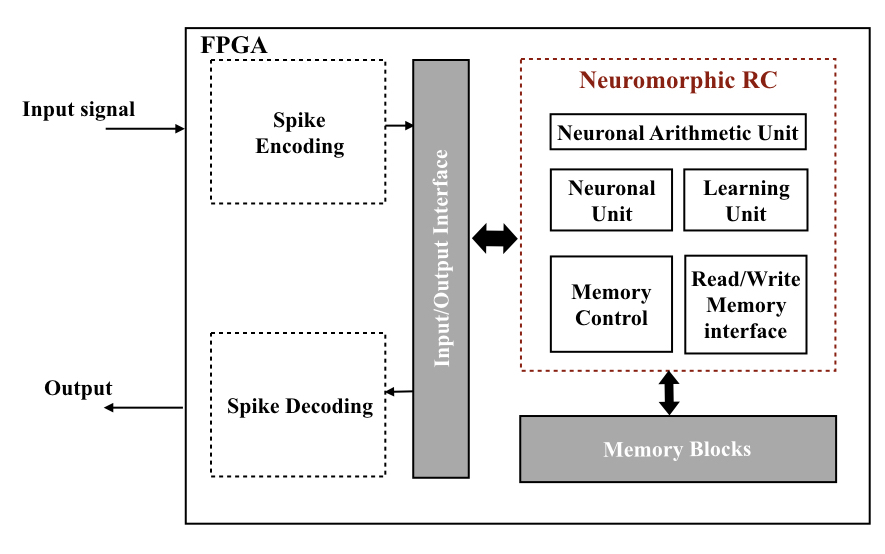}
		\par\end{centering}
	\caption{A top-level schematic of an FPGA-based neuromporphic reservoir computer.}
	\label{fig: Fig1}
\end{figure}

As illustrated in Figure \ref{fig: Fig1}, an FPGA-based reservoir computer consists of neuronal units, a neuronal arithmetic section, input/output encoding/decoding components, an on-chip learning algorithm, memory control, numerous memory blocks such as RAMs and/or memristors wherein the neuronal and synaptic information is stored, and a Read/Write interface to access to these memory blocks. Realizing spiking reservoir computing on such substrates entails tackling a number of critical issues related to spike coding, memory organization, parallel processing, on-chip learning and tradeoffs between area, precision and power overheads. Moreover, in real-world applications such as speech recognition and biosignal processing, spiking dynamics might be inherently faster than real-time performance. In this section, an overview of experimental studies is provided to illustrate how these challenges have been so far addressed in the literature.

In the first digital implementation of spiking RC, \cite{schrauwen2007compact} suggested storing the neuronal information in RAM, exploiting non-plastic exponential synapses and performing neuronal and synaptic operations serially. In this clock-based simulation, each time step consists of several operations such as adding weights to the membrane or synapse accumulators, adding the synapse accumulators to the membrane’s, decaying these accumulators, threshold detection and membrane reset. On a real-time speech processing task, this study shows that in contrast to ``serial processing, parallel arithmetic" \cite{upegui2005fpga} and ``parallel processing, serial arithmetic" \cite{schrauwen2006parallel}, serial implementations of both arithmetic operations that define the dynamics of the neuron model and processing operations associated with synaptic dynamics yield to slower operations and low hardware costs. Although this architecture can be considered as the first compact digital neuromorphic RC system to solve a speech recognition task, the advantage of distributed computing is not explored in this framework. Moreover, for larger networks and in more sophisticated tasks, a sequential processor -- which calculates every neuronal unit separately to simulate a single time step of the network -- does not seem to be efficient. Stochastic arithmetic was, therefore, exploited to parallelize the calculations and obtain a considerable speedup \cite{verstraeten2005reservoir}.

More biologically plausible models of spiking neurons (e.g., Hodgkin--Huxley, Morris--Lecar and Izhikevich models \cite{gerstner2002spiking}) are too sophisticated to be efficiently implemented on hardware and have many parameters which need to be tuned. On the other hand, the simple hardware-friendly spiking neuron models, such as leaky-integrate-and-fire (LIF) \cite{gerstner2002spiking}, often have a hard thresholding that makes supervised training difficult in spiking neural networks (SNNs). In spiking reservoir computing, however, the training of the readout mechanism amounts only to solving a linear regression problem, where the target output is a trainable linear combination of the neural signals within the reservoir. Linear regression is easily solved by standard linear algebra algorithms when arbitrary real-valued combination weights are admitted.
However, for on-chip learning, the weights will be physically realized by states of electronic synapses, which currently can be reliably set only to a very
small number of discrete values. It has been recently proved that computing optimal discrete readout weights in
reservoir computing is NP-hard and approximate (or heuristic) methods must be exploited to obtain high-quality solutions in reasonable time for practical uses \cite{hadaeghi2019computing}. The spike-based learning algorithm proposed by \cite{zhang2015digital} is an example of such approximate solutions for FPGA implementations. In contrast to offline learning methods, the proposed online learning rule avoids any intermediate data storage. Besides, through this abstract learning process, each synaptic weight is adjusted based on only the firing activities of the corresponding pre- and post-synaptic neurons, independent of the global communications across the neural network. In a speech recognition task, it has been shown that due to this locality, the overhead of hardware implementation (e.g., synaptic and membrane voltage precision) can be reduced without drastic effects on its performance \cite{zhang2015digital}. This biological-inspired supervised learning algorithm was later exploited in developing an RC-based general-purpose neuromorphic architecture where training the task-specific output neurons is integrated into the reconfigurable FPGA platform \cite{wang2015general}. The effects of catastrophic failures such as broken synapses, dead neurons, random errors as well as errors in arithmetic operations (e.g., comparison, adding and shifting) in similar architecture were addressed in \cite{jin2017performance} where the reservoir consists of excitatory and inhibitory LIF neurons that are randomly connected with non-plastic synapses governed by second-order response functions. The simulation results suggest that at least 8-bit resolution is needed for the efficacy of plastic synapses if the spike-based learning algorithm proposed by \cite{zhang2015digital} is applied for training the readout weights. The recognition performance only degrades slightly when the precision of fixed synaptic weights and membrane voltage for reservoir neurons are reduced down to 6 bits. 

To realize the on-chip learning on digital systems with extremely low-bit precision (binary values), \cite{jin2016sso} suggested an online pruning algorithm based on variances of firing activities to sparsify the readout connections. The cores to this reconfiguration scheme are STDP learning rule, firing activity monitoring, and variance estimation. The readout synapses projected from low-variance reservoir neurons are, then, powered off to save energy.
To reduce the energy consumption of the hardware in FPGA, the reservoir computer developed by \cite{wang2017energy,wang2016liquid} utilizes firing activity based power gating by turning off the neurons that seldom fire for a particular benchmark, and applies approximate arithmetic computing \cite{shao2015array} to speed up the runtime in a speech recognition task. 

Exploiting the
spatial locality in a reservoir consisting of excitatory and inhibitory LIF neurons, a fully parallel design approach was also presented for real-time biomedical signal processing in \cite{polepalli2016digital,polepalli2016reconfigurable}. In this design, the memory blocks were replaced by distributed memory to circumvent the long access time due to wire delay. Being inspired by new  discoveries in neuroscience, \cite{smith2017novel} developed a spiking temporal processing unit (STPU) to efficiently implement more biologically plausible synaptic response functions in digital architectures. STPU offers a local temporal memory buffer including an arbitrary number of memory cells to model delayed synaptic transmission between neurons. This allows multiple connections between neurons with different synaptic latency. Utilizing excitatory and inhibitory LIF neurons with different time scales and exploiting the second-order response function to model synaptic transmission, the effects of the input signal in a spoke digit recognition task will only slowly ``wash out'' over time, enabling the reservoir to provide sufficient dynamical short-term memory capacity. In order to create a similar short-term memory on a reconfigurable digital platform with extremely low bit resolutions, an STDP mechanism was proposed by \cite{jin2016sso} for on-chip reservoir tuning. Applying this local learning rule to the reservoir synapses together with a data-driven binary quantization algorithm creates sparse connectivities within the reservoir in a self-organized fashion, leading to significant energy reduction. Stochastic activity-based STDP approach \cite{jin2016ap}, structural plasticity-based mechanism \cite{roy2016online}, and correlation-based neuron gating rule \cite{liu2018online} have also been introduced for efficient low-resolution tuning of the reservoir in hardware. Neural processes, however, require a large number of memory resources for storing various parameters, such as synaptic weights and internal neural states, and lead to a heavy clock distribution loading and a significant power dissipation. To tackle this problem, \cite{liu2018online} suggested partitioning the memory elements inside each neuron that are activated at different phases of neural processing. This leads to an activity-based clock gating mechanism with a granularity of a partitioned memory group inside each neuron. 

Another technique for power reduction is incorporating memristive crossbars into digital neuromorphic hardware to perform synaptic operations in on-chip/online learning and to store the synaptic weights in offline learning \cite{soures2017robustness, moon2019temporal, midya2019reservoir}. In general, a two terminal
memristor device can be programmed by applying a
large enough potential difference across its terminals, where the state change of the device is dependent on the magnitude, duration, and polarity of the potential difference. The device resistance states, representing synaptic weights, will vary between a high resistance state (HRS) or low resistance state (LRS), depending on the polarity of the voltage applied. At a system level, various sources of noise (e.g., random telegraph, thermal noise, and 1/f noise) arise from non-ideal behavior in memristive devices and distort the process of reading from the synapses and updating the synaptic weights. Given theoretical models for these stochastic noise processes,  the effects of different manifestations of memristor read and write noise on the accuracy of neuromorphic RC in a classification task were investigated in \cite{soures2017robustness}. Other hardware feasible RC structure with memristor double crossbar array have also been proposed in \cite{hassan2017hardware, moon2019temporal, midya2019reservoir, wlazlak2020neuromorphic} and were tested on a real-time time series prediction/classification tasks. These studies not only confirm that reservoir computing can properly cope with low-precision environments and noisy data, but they also experimentally demonstrate how RC benefits from memristor device variation to secure a more random heterogeneous weight distribution leading to more diverse response for similar inputs. Although memristive crossbars are area and energy efficient, the digital circuitry to control the read/write logic to the crossbars is extremely power hungry, thus preventing the use of memristors in large-scale memory systems.

In RC literature, it has also been shown that in time series prediction tasks, a cyclic reservoir --where neurons are distributed on a ring and connected with the same synaptic weights to produce the cyclic rotations of the input vector-- is able to operate with an efficiency comparable to the standard RC models \cite{rodan2011minimum,appeltant2011information}. In order to minimize the required hardware resources, therefore, single cyclic reservoir systems with stochastic spiking neurons were implemented for real-time time series prediction and classification tasks \cite{alomar2016stochastic, wlazlak2020neuromorphic}.The architectures show high level of scalability and prediction performance comparable to the software simulations. In non-spiking reservoir computing implementations on reconfigurable digital systems, the reservoir has sometimes been configured as a ring topology and the readout weights are trained through gradient descent algorithm or ridge regression \cite{antonik2015fpga,yi2016fpga,alomar2016fpga}.


\section{RC on Analog Neuromorphic Microchips}
\begin{figure} [tp]
	\begin{centering}
		\includegraphics[width=1.0\columnwidth]{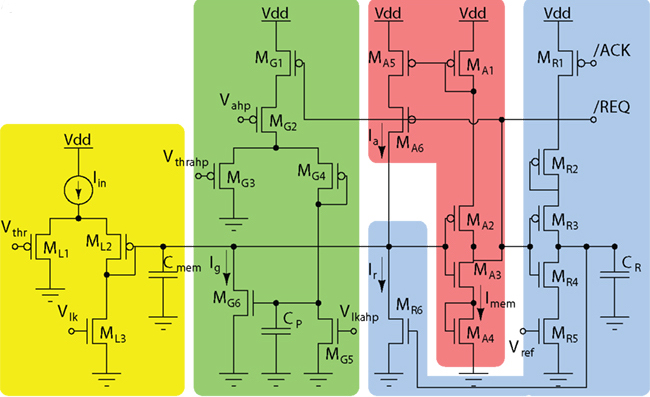}
		\par\end{centering}
	\caption{A schematic of (DPI) neuron circuit (reproduced from \cite{indiveri2011neuromorphic}).}
	\label{fig: Fig2}
\end{figure}

Digital platforms, tools and simulators offer robust and practical solutions to a wide range of engineering problems and provide convenient approaches to explore the quantitative behavior of neural networks. However, they do not seem to be ideal substrates for simulating the detailed large-scale models of neural systems where  density, energy efficiency, and resilience are important. Besides, the observation that the brain operates on analog principles and relies on the inherent physical characteristics of the neuronal system for computation motivated the investigations in the field of neuromorphic engineering. Following the pioneering work conducted by Carver Mead \cite{mead2012analog} -- therein biologically inspired electronic sensors were integrated with analog circuits and an address-event-based asynchronous, continuous time communications protocol was introduced -- over the last decade, mixed analog/digital and purely analog very large scale integration (VLSI) circuits have been fabricated to emulate the electrophysiological behavior of biological neurons and synapses and to offer a medium in which neuronal networks can be presented directly in hardware rather than simply simulated on a general purpose computers \cite{indiveri2011neuromorphic}. However, the fact that such platforms provide only a qualitative approximation to the exact performance of digitally simulated neural systems may preclude them from being the ideal substrate for solving engineering problems and achieving machine learning tasks where detailed quantitative investigations are essential. In addition to the low resolution, realizing global asynchrony and dealing with noisy and unreliable components seem to be formidable technical hurdles. An architectural solution can be partly provided by the reservoir computing due to its bio-inspired principle, robustness to the imperfect substrate and the fact that only readout weights need to be trained \cite{schurmann2005edge}.  
Similar to other analog and digital computing chips, transistors are the building blocks of analog neuromorphic devices. It has been experimentally shown that in the ‘‘subthreshold’’ region of operation, the current-voltage characteristic curve of the transistor is exponential and analogous to the exponential dependence of active populations of voltage-gated ionic channels as a function of the potential across the membrane of a neuron. This similarity has paved the way towards the fabrication of compact analog circuits that implement electronic models of
voltage-sensitive conductance-based neurons and conductance-based synapses as well as computational circuitry to perform logarithmic functions, amplification, thresholding, multiplication, inhibition, and winner-take-all selection \cite{liu2010neuromorphic,donahue2015design}. An example of biophysically realistic neural electronic circuits is depicted in Figure \ref{fig: Fig2}. This differential pair integrator (DPI) neuron circuit consists of four components: 1) the input DPI filter ($M_{L1}$ -- $M_{L3}$) including the integrating membrane capacitor $C_{mem}$, models the neuron’s leak conductance which produces exponential sub-threshold dynamics in response to constant input currents, 2) a spike event generating amplifier ($M_{A1}$ -- $M_{A6}$) together with a positive-feedback circuit represent both sodium channel activation and inactivation dynamics, 3) a spike reset circuit with address event representation (AER) handshaking signals and refractory period functionality ($M_{R1}$ -- $M_{R6}$) emulates the potassium conductance functionality and 4) a spike-frequency adaptation mechanism implemented by an additional DPI filter ($M_{G1}$ -- $M_{G6}$) produces an after hyper-polarizing current proportional to the neuron’s mean firing rate. See \cite{indiveri2011neuromorphic} for an insightful review of different design methodologies used for silicon neuron fabrication.

In conjunction with circuitry that operates in subthreshold mode, exploiting memristive devices to model synaptic dynamics is a common approach in analog neuromorphic systems for power efficiency and performance boosting purposes \cite{indiveri2015memory,avizienis2012neuromorphic}. In connection with the fully analog implementation of RC, it has been shown that by creating a functional reservoir consisting of numerous neurons connected through atomic nano-switches, the functional characteristics required for implementing biologically inspired computational methodologies in a synthetic experimental system can be displayed \cite{avizienis2012neuromorphic}. A memristor crossbar structure has also been fabricated to connect the analog non-spiking reservoir neurons -- which are distributed on a ring topology -- to the output node \cite{donahue2015design,merkel2014memristive}. In this CMOS compatible crossbar, multiple weight states are achieved using multiple bi-stable memristors (with
only two addressable resistance states). The structural simplicity of this architecture paves the way to independent control of each synaptic element. However, the discrete nature of memristive weights places a limit
on the accuracy of reservoir computing; thus the number of weight states per synapse, that are required for satisfactory accuracy, must be determined beforehand. Besides, appropriate learning algorithms for on-chip training purpose have not yet been implemented for fully analog reservoir computing.
Another significant design challenge associated with memristor devices is cycle-to-cycle variability in the resistance values even within a single memristor. With the same reservoir structure, therefore, \cite{yang2016investigations} showed by connecting memristor devices in series or
parallel, a staircase memristor model could be constructed which not only has a delayed-switching
effect between several somewhat stable resistance levels, but also can provide more reliable state values if a specific resistance level is required. This model of synaptic delay is particularly relevant for time delay reservoir methodologies \cite{appeltant2011information}.

From the information processing point of view, the ultimate aim of neuromorphic systems is to carry out neural computation in an energy-efficient way. However, firstly, the quantities relevant to the computation have to be expressed in terms of the spikes that spiking neurons communicate with. One of the key components in both digital and analog neuromorphic reservoir computers is, therefore, a neural encoder which transforms the input signals into the spike trains. Although the nature of the neural code (or neural codes) is an unresolved topic of research in neuroscience, based on what is known from biology, a number of neural information encodings have been proposed \cite{gruning2014spiking}. Among them, hardware prototypes of rate encoding, temporal encoding, inter-spike interval encoding, and latency (or rank order) encoding schemes have been implemented, mostly on digital platforms \cite{yi2016fpga,yang2016investigations,bichler2012extraction,zhao2016energy}. Digital configurations, however, requires large hardware overheads associated with analog-to-digital converters, operational amplifiers, digital buffers, and electronic synchronizers and will increase the cost of implementation. Particularly, in analog reservoir computers, fabricating a fully analog encoding spike generator is of crucial importance both to speed the process up and to optimize the required energy and hardware costs. Examples of such generators have been proposed, mainly, for analog implementation of delayed feedback reservoirs \cite{zhao2016energy,li2017analog}.


\section{RC on Mixed Digital/Analog Neuromorphic Systems}
The majority of analog neuromorphic systems designed to solve machine learning/signal processing problems tend to rely on digital components, for instance, for pre-processing, encoding spike generation, storing the synaptic weights values and on-chip programmability/learning \cite{murray1988asynchronous,azghadi2013programmable,briiderle2010simulator}. Inter-chip and chip-to-chip communications are also primarily digital in some analog neuromorphic platforms \cite{murray1988asynchronous,aamir2016highly,chicca2006modeling}. Mixed analog/digital architectures are, therefore, very common in neuromorphic systems. These electronic chips often provide a universal substrate with an extensive configuration space to emulate different types of neurons, synapses, and network typologies. On the hardware developed in \cite{pfeil2013six}, for instance, together with five other neural network models, a spiking reservoir model consisting of excitatory and inhibitory populations, and a readout tempotron neuron was implemented and trained to classify spike train segments in a continuous data stream. The hardware offers on-chip spike-based learning mechanisms to compensate for fixed-pattern noise. 

\begin{figure} [tp]
	\begin{centering}
		\includegraphics[width=1.0\columnwidth]{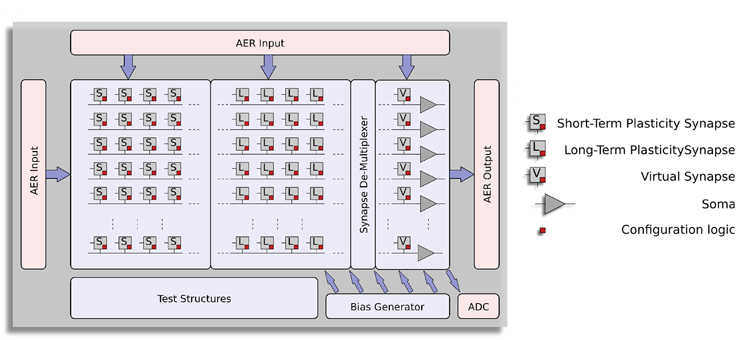}
		\par\end{centering}
	\caption{Architecture of ROLLS neuromorphic processor (reproduced from \cite{qiao2015reconfigurable}).}
	\label{fig: Fig3}
\end{figure}

In the context of hardware reservoir computing and inspired by the nonlinear properties of dendrites in biological neurons, \cite{roy2014liquid} proposed a readout learning mechanism which returns binary values for synaptic weights such that the ``choice" of connectivity can be implemented in a mixed analog/digital platform with the address
event representation (AER) protocols where the connection matrix is stored in digital memory. Relying on the same communication protocol and exploiting the reconfigurable online learning spiking neuromorphic processor (ROLLS chip) introduced in \cite{qiao2015reconfigurable}, \cite{corradi2015neuromorphic} designed and fabricated a reservoir computer to detect spike-patterns in bio-potential and local field potential (LFP) recordings. The ROLLS neuromorphic processor (Figure \ref{fig: Fig3}) contains 256 adaptive exponential integrate-and-fire (spiking) neurons implemented with mixed signal analog/digital circuits. The neurons
are connected to an array of $256 \times 256$ learning synapse circuits for modeling long-term plasticity mechanisms, an array of $256 \times 256$ programmable synapses with short-term plasticity circuits and $256 \times 2$ row of ``virtual synapses" for modeling excitatory and inhibitory synapses that have shared synaptic weights and time constants. The reservoir in this model comprises 128 randomly connected spiking neurons that receive spike event inputs from a neural recording system and aim to enhance the temporal properties of input patterns. The readout layer is afterwards trained by applying a spike-based Hebbian-like learning rule previously proposed in \cite{brader2007learning}. The results of this study suggest that the device mismatch and the limited precision of the analog circuits result in a diversity of neuronal responses that might be beneficial in a population coding scheme within the reservoir computing framework.

Later, in another implementation of reservoir computing on mixed analog/digital substrates endowed with learning abilities, the neuronal circuits were chosen to be analog for power and area efficiency considerations and an on-chip learning mechanism based on recursive least squares (RLS) algorithm was implemented on an FPGA platform \cite{yi2016fpga}. The proposed architecture was then applied to modeling a Multiple-Input Multiple-Output Orthogonal Frequency Division Multiplexing (MIMO-OFDM) system.
A completely on-chip feasible design for RC has also been proposed in \cite{kudithipudi2016design}, where non-spiking reservoir neurons are distributed in a hybrid topology and memristive nano synapses were used in the output or regression layer. The proposed hybrid topology consists of a ring RC and a center neuron connected to all the reservoir neurons which transmit the information about the state of the whole reservoir to each neuron. To fabricate non-spiking silicon neurons, current-mode differential amplifiers operating in their subthreshold regime have been used to emulate the hyperbolic tangent rate model behavior. For epileptic seizure detection and prosthetic finger
control, it has been shown that a random distribution of weights in input-to-reservoir and reservoir synapses can be obtained by employing mismatches in transistors threshold voltages to design subthreshold bipolar synapses.

In order to create a functional spiking reservoir on the Dynap-se board \cite{moradi2017scalable}, a ``Reservoir Transfer Method" was also proposed to transfer the functional ESN-type reservoir into a reservoir with analog, unclocked, spiking neurons \cite{he2018EMBS}. The Dynap-se board is a multicore neuromorphic processor chip that employs hybrid analog/digital circuits for emulating synapse and neuron dynamics together with asynchronous digital circuits for managing the address-event traffic. It offers 4000 adaptive exponential integrate-and-fire (spiking) neurons implemented with mixed signal analog/digital circuits with  $64/4 k$ fan-in/out. From the computational point of view, implementing efficient algorithms on this neuromorphic system encounters general problems such as bit resolution, device mismatch, uncharacterized neural models, unavailable state variables, and physical system noise. Realizing the reservoir computing on this substrate, additionally, leads to an important problem associated with fine-tuning the reservoir to obtain the memory span required for a specific machine learning problem. The reservoir transfer method proposes a theoretical framework to learn ternary weights in a reservoir of spiking neurons by transferring features from a well-tuned echo state network simulated on software \cite{he2018EMBS}. Empirical results from an ECG signal monitoring task showed that this reservoir with ternary weights is able to not only integrate information over a time span longer than the timescale of individual neurons but also function as an information processing medium with performance close to a standard, high precision, deterministic, non-spiking ESN \cite{he2018EMBS}.

\section{Conclusion}
Reservoir computing appears to be a particularly widespread
and versatile approach to harness unconventional nonlinear physical phenomena into useful computations. Spike-based neuromorphic microchips, on the other hand, promise one or two orders of magnitude less energy consumption than traditional digital microchips. Implementation of reservoir computing methodologies on neuromorphic hardware, therefore, has been an attractive practice in neuromorphic system design. Here, a review of experimental studies was provided to illustrate the progress in this area and to address the computational bottlenecks which arise from specific hardware implementations. Moreover, to deal with challenges of computation on such unconventional substrates, several lines of potential solutions are presented based on advances in other computational approaches in machine learning.

\section*{Acknowledgments} This work was supported by European H2020 collaborative project NeuRAM3 [grant number 687299]. I would also like to thank Herbert Jaeger, who provided insight and expertise that greatly assisted this research.

\bibliographystyle{unsrt}
\bibliography{RcBookChapter}
\end{document}